# Evolution of self-gravitating fluid spheres involving ghost stars

Luis Herrera
Instituto Universitario de Física Fundamental y Matemáticas,
Universidad de Salamanca, Salamanca 37007,
Spain *
Alicia Di Prisco
Escuela de Física. Facultad de Ciencias.
Universidad Central de Venezuela. Caracas, Venezuela.
Justo Ospino
Departamento de Matemática Aplicada
Instituto Universitario de Física Fundamental y Matemáticas,
Universidad de Salamanca,
Salamanca 37007, Spain

**Abstract**

Exact solutions are presented which describe, either the evolution of fluid distributions corresponding to a ghost star (vanishing total mass), or describing the evolution of fluid distributions which attain the ghost star status at some point of their lives. The first two solutions correspond to the former case, they admit a conformal Killing vector (CKV) and describe the adiabatic evolution of a ghost star. Other two solutions corresponding to the latter case are found, which describe evolving fluid spheres absorbing energy from the outside, leading to a vanishing total mass at some point of their evolution. In this case the fluid is assumed to be expansion–free. In all four solutions the condition of vanishing complexity factor was imposed. The physical implications of the results, are discussed

---

*E-mail address: lherrera@usal.es

# 1 Introduction

In a recent paper [1] we have explored the possibility of finding fluid distributions of anisotropic fluids whose total mass is zero. Such configurations which we named ghost stars, are characterized by the presence of negative energy-density within the distribution, as consequence of which the total mass vanishes. The solutions presented in [1] are static, and may be regarded as either the final or the initial state of a dynamical process.

At this point the following remarks are in order

- As mentioned above the vanishing total mass of ghost stars is the result of the appearance of negative energy-density within some regions of the fluid distribution. Therefore such objects should be clearly differentiated from "evaporated" stars which have radiated all its energy away leading to the vanishing of its energy-density.

- The observational data collected from compact objects relies almost exclusively on two physical phenomena, namely: surface gravitational redshift and the influence of the gravitational field of the object on light rays passing close to it (its shadow). Both of which of course depend on the gravitational surface potential of the object (vanishing for any ghost star), which illustrates the potential observational consequences of such objects (see further comments on this issue in the last section).

In this work we endeavor to describe the evolution of fluid distributions describing the following possible scenarios

- The adiabatic evolution of a ghost star. The total mass remains zero all along the evolution.

- The non–adiabatic evolution of fluid distribution reaching at some point a zero total mass (with non-vanishing energy-density).

For the first scenario we obtain two solutions admitting a CKV. One of them corresponds to a CKV orthogonal to the four–velocity vector of the fluid, whereas the other admits a CKV parallel to the four–velocity.

The motivation behind the admittance of CKV is provided by the fact that it generalizes the well known concept of self–similarity, which play a very important role in classical hydrodynamics.

This explains the great interest aroused by this kind of symmetry (for very recent developments on this issue see, for example, [2]-[12] and references therein).

This first couple of solutions represents fluid distributions contracting from arbitrarily large areal radius to compact fluid distribution with finite areal radius, or fluid distributions expanding from some finite value of the areal radius to infinity, always satisfying the vanishing total mass condition. The Darmois conditions at the boundary are satisfied, and as expected the energy-density is negative in some regions of the fluid distribution.

In the second scenario both solutions satisfy the vanishing expansion scalar condition. It is worth recalling that such a condition implies the existence of a cavity around the center, thereby helping in the modeling of voids observed at cosmological scales.

Newest results regarding expansion–free fluids may be found in [13]- [24] and references therein.

The two solutions satisfying the expansion–free condition represent fluid distribution contracting from arbitrarily large configurations to a singularity. They evolve absorbing energy from the outside, and at some point of their evolution their total mass vanishes. Thus the solutions pass through a ghost star state.

The presented solutions satisfy additional conditions which allow the full integration of field equations. These restrictions are the vanishing complexity factor condition as defined in [25, 26] and the quasi–homologous evolution [27]. At any rate it is worth recalling that none of the above conditions is intrinsically related to the ghost star condition, and are assumed with the sole purpose to allow analytical integration of the Einstein equations.

A discussion on all the presented models is brought out in the last section.

## 2 The relevant equations and variables

The system we are dealing with consists in a spherically symmetric distribution of collapsing fluid, bounded by a spherical surface $\Sigma$. The fluid is assumed to be locally anisotropic (principal stresses unequal) and undergoing dissipation in the form of heat flow (diffusion approximation). We shall proceed now to summarize the definitions and main equations required for describing spherically symmetric dissipative fluids. A detailed description may be found in [27].

## 2.1 The metric, the energy–momentum tensor, the kinematic variables and the mass function

Choosing comoving coordinates, the general interior metric can be written as

$$ds^2 = -A^2 dt^2 + B^2 dr^2 + R^2(d\theta^2 + \sin^2\theta d\phi^2). \tag{1}$$

The energy momentum tensor in the canonical form, reads

$$T_{\alpha\beta} = \mu V_\alpha V_\beta + P h_{\alpha\beta} + \Pi_{\alpha\beta} + q(V_\alpha K_\beta + K_\alpha V_\beta), \tag{2}$$

with

$$P = \frac{P_r + 2P_\perp}{3}, \qquad h_{\alpha\beta} = g_{\alpha\beta} + V_\alpha V_\beta,$$

$$\Pi_{\alpha\beta} = \Pi\left(K_\alpha K_\beta - \frac{1}{3}h_{\alpha\beta}\right), \qquad \Pi = P_r - P_\perp,$$

where $\mu$ is the energy-density, $P_r$ the radial pressure, $P_\perp$ the tangential pressure, $q^\alpha = qK^\alpha$ the heat flux, $V^\alpha$ the four–velocity of the fluid, and $K^\alpha$ a unit four–vector along the radial direction. For comoving observers, we have

$$V^\alpha = A^{-1}\delta^\alpha_0, \quad K^\alpha = B^{-1}\delta^\alpha_1, \tag{3}$$

satisfying

$$V^\alpha V_\alpha = -1, \quad V^\alpha q_\alpha = 0, \quad K^\alpha K_\alpha = 1, \quad K^\alpha V_\alpha = 0. \tag{4}$$

Both bulk and shear viscosity, as well as dissipation in the free streaming approximation can be trivially absorbed in $P_\perp, \mu, P_r$ and $q$.

The Einstein equations for (1) and (2), are explicitly written in Appendix A.

The acceleration $a_\alpha$ and the expansion $\Theta$ of the fluid are given by

$$a_\alpha = V_{\alpha;\beta} V^\beta, \quad \Theta = V^\alpha{}_{;\alpha}, \tag{5}$$

and its shear $\sigma_{\alpha\beta}$ by

$$\sigma_{\alpha\beta} = V_{(\alpha;\beta)} + a_{(\alpha} V_{\beta)} - \frac{1}{3}\Theta h_{\alpha\beta}. \tag{6}$$

From the equations above we have for the four–acceleration and its scalar $a$,

$$a_\alpha = aK_\alpha, \quad a = \frac{A'}{AB}, \tag{7}$$

and for the expansion
$$\Theta = \frac{1}{A}\left(\frac{\dot{B}}{B} + 2\frac{\dot{R}}{R}\right), \tag{8}$$
while for the shear we obtain
$$\sigma^{\alpha\beta}\sigma_{\alpha\beta} = \frac{2}{3}\sigma^2, \tag{9}$$
where
$$\sigma = \frac{1}{A}\left(\frac{\dot{B}}{B} - \frac{\dot{R}}{R}\right), \tag{10}$$
in the above prime stands for $r$ differentiation and the dot stands for differentiation with respect to $t$.

Next, the mass function $m(t,r)$ reads [28]
$$m = \frac{R^3}{2}R_{23}{}^{23} = \frac{R}{2}\left[\left(\frac{\dot{R}}{A}\right)^2 - \left(\frac{R'}{B}\right)^2 + 1\right]. \tag{11}$$

Introducing the proper time derivative $D_T$ given by
$$D_T = \frac{1}{A}\frac{\partial}{\partial t}, \tag{12}$$
we can define the velocity $U$ of the collapsing fluid as the variation of the areal radius with respect to proper time, i.e.
$$U = D_T R, \tag{13}$$
where $R$ defines the areal radius of a spherical surface inside the fluid distribution (as measured from its area).

Then (11) can be rewritten as
$$E \equiv \frac{R'}{B} = \left(1 + U^2 - \frac{2m}{R}\right)^{1/2}. \tag{14}$$

An alternative expression for $m$ may be found using field equations, it reads
$$m = 4\pi \int_0^r \left(\mu + q\frac{U}{E}\right) R^2 R' dr, \tag{15}$$
satisfying the regular condition $m(t,0) = 0$.

From the above equation it follows that, since $R' > 0$ in order to avoid shell crossing singularities, the vanishing total mass condition ($m(r_\Sigma) = 0$) requires that the "effective energy- density" ($\mu + q\frac{U}{E}$) should be either zero (the trivial case), or changes its sign within the fluid distribution (ghost star).

## 2.2 The complexity factor

The complexity factor is scalar that measures the degree of complexity of a given fluid distribution [25, 26]. It is identified with the scalar function $Y_{TF}$ which defines the trace–free part of the electric Riemann tensor (see [25, 26] for details).

Using field equations we obtain

$$Y_{TF} = -8\pi\Pi + \frac{4\pi}{R^3}\int_0^r R^3\left(\mu' - 3q\frac{UB}{R}\right)dr. \quad (16)$$

In terms of the metric functions the scalar $Y_{TF}$ reads

$$Y_{TF} = \frac{1}{A^2}\left[\frac{\ddot{R}}{R} - \frac{\ddot{B}}{B} + \frac{\dot{A}}{A}\left(\frac{\dot{B}}{B} - \frac{\dot{R}}{R}\right)\right]$$
$$+ \frac{1}{B^2}\left[\frac{A''}{A} - \frac{A'}{A}\left(\frac{B'}{B} + \frac{R'}{R}\right)\right]. \quad (17)$$

## 2.3 The homologous and quasi–homologous conditions

In the dynamic case, the discussion about the complexity of a fluid distribution involves not only the complexity factor which describes the complexity of the structure of the fluid, but also the complexity of the pattern of evolution.

Following previous works [26, 27] we shall consider two specific modes of evolution as the most suitable candidates to describe the simplest pattern of evolution. These are, the homologous evolution (H) [26] characterized by

$$U = \tilde{a}(t)R, \qquad \tilde{a}(t) \equiv \frac{U_\Sigma}{R_\Sigma}, \quad (18)$$

and

$$\frac{R_1}{R_2} = \text{constant}, \quad (19)$$

where $R_1$ and $R_2$ denote the areal radii of two concentric shells $(1, 2)$ described by $r = r_1 = $ constant, and $r = r_2 = $ constant, respectively.

A somehow softer condition is represented by the quasi–homologous condition (QH) [27], which only requires the fulfillment of (18).

It can be shown, using the field equations (see [26, 27] for details) that (18) implies

$$\frac{4\pi}{R'}Bq + \frac{\sigma}{R} = 0. \quad (20)$$

Thus, H condition implies (19) and (20), whereas QH condition only implies (20).

## 2.4 The exterior spacetime and junction conditions

Since our fluid distribution is bounded we assume that outside $\Sigma$ the space–time is described by Vaidya metric which reads.

$$ds^2 = -\left[1 - \frac{2M(v)}{\mathbf{r}}\right]dv^2 - 2d\mathbf{r}dv + \mathbf{r}^2(d\theta^2 + \sin^2\theta d\phi^2), \qquad (21)$$

where $M(v)$ denotes the total mass, and $v$ is the retarded time.

The smooth matching of the full nonadiabatic sphere to the Vaidya spacetime, on the surface $r = r_\Sigma =$ constant, requires the fulfillment of the Darmois conditions, i.e. the continuity of the first and second fundamental forms across $\Sigma$ (see [29] and references therein for details), which implies

$$m(t,r) \stackrel{\Sigma}{=} M(v), \qquad (22)$$

and

$$q \stackrel{\Sigma}{=} P_r, \qquad (23)$$

where $\stackrel{\Sigma}{=}$ means that both sides of the equation are evaluated on $\Sigma$.

When Darmois conditions are not satisfied the boundary surface is a thin shell.

## 2.5 The transport equation

In the case of non–adiabatic evolution we have to resort to some transport equation to describe the evolution and spatial distribution of the temperature. Thus for example within the context of the Israel– Stewart theory [30, 31, 32] the transport equation for the heat flux reads

$$\begin{aligned}\tau h^{\alpha\beta}V^\gamma q_{\beta;\gamma} + q^\alpha &= -\kappa h^{\alpha\beta}\left(T_{,\beta} + Ta_\beta\right) \\ &\quad - \frac{1}{2}\kappa T^2 \left(\frac{\tau V^\beta}{\kappa T^2}\right)_{;\beta} q^\alpha,\end{aligned} \qquad (24)$$

where $\kappa$ denotes the thermal conductivity, and $T$ and $\tau$ denote temperature and relaxation time respectively.

In the spherically symmetric case under consideration, the transport equation has only one independent component which may be obtained from (24) by contracting with the unit spacelike vector $K^\alpha$, we get

$$\tau V^\alpha q_{,\alpha} + q = -\kappa \left(K^\alpha T_{,\alpha} + Ta\right) - \frac{1}{2}\kappa T^2 \left(\frac{\tau V^\alpha}{\kappa T^2}\right)_{;\alpha} q. \tag{25}$$

Sometimes it is possible to simplify the equation above, in the so called truncated transport equation, when the last term in (24) may be neglected [33], producing

$$\tau V^\alpha q_{,\alpha} + q = -\kappa \left(K^\alpha T_{,\alpha} + Ta\right). \tag{26}$$

## 3 Exact solutions

We shall now proceed to present exact analytical solutions describing two different scenarios,

- the evolution of a ghost star keeping its nature ($M = 0$) all along its evolution

- the evolution of a compact object (not a ghost star) attaining the ghost star condition at some moment of its life.

We shall consider, both, the non–dissipative and the dissipative case.

In order to specify our models we need to impose some further restrictions. In this work such restriction will be

- The admittance of a conformal killing vector (CKV),

   or

- The expansion–free condition.

In some cases the above conditions have to be complemented with additional restrictions such as

- The vanishing complexity factor condition.

- The homologous or the quasi–homologous approximation.

## 3.1 Solutions admitting a CKV

In this subsection we shall consider spacetimes satisfying the equation

$$\mathcal{L}_\chi g_{\alpha\beta} = 2\psi g_{\alpha\beta}, \tag{27}$$

where $\mathcal{L}_\chi$ denotes the Lie derivative with respect to the vector field $\chi$, which unless specified otherwise, has the general form

$$\chi = \xi(t,r,)\partial_t + \lambda(t,r,)\partial_r, \tag{28}$$

and $\psi$ in principle is a function of $t$ and $r$. The case $\psi = constant$ corresponds to a homothetic Killing vector (HKV). The solutions described here are particular cases of solutions found in [8].

We shall consider two possible subclasses, both of which describe non–dissipative evolution

- $\chi^\alpha$ orthogonal to $V^\alpha$,
- $\chi^\alpha$ parallel to $V^\alpha$.

In the first case ($\chi^\alpha$ orthogonal to $V^\alpha$), we shall obtain from the matching conditions, the QH condition and the vanishing complexity factor condition, with $M = 0$, solution $I$.

In the second case ($\chi^\alpha$ parallel to $V^\alpha$), we shall obtain from the matching conditions and the vanishing complexity factor condition, solution $II$.

Let us start by considering the case $\chi^\alpha$ orthogonal to $V^\alpha$, and $q = 0$.

### 3.1.1 Solution I: $\chi_\alpha V^\alpha = q = M = 0$.

In this case we obtain from (27) (see [8] for details)

$$A = \alpha R = \frac{F(t)}{f(t) + g(r)}, \quad B = \frac{1}{f(t) + g(r)}, \tag{29}$$

where $f$ and $g$ are two arbitrary functions of their arguments and $\alpha$ is a unit constant with dimensions of $[\frac{1}{r}]$.

Thus any model is determined up to three arbitrary functions $F(t), f(t), g(r)$, in terms of which the field equations read

$$8\pi\mu = \frac{(f+g)^2}{F^2}\left[\frac{\dot{F}^2}{F^2} - \frac{4\dot{F}\dot{f}}{F(f+g)} + \frac{3\dot{f}^2}{(f+g)^2}\right]$$
$$+ 2g''(f+g) - 3g'^2 + \frac{\alpha^2(f+g)^2}{F^2}, \tag{30}$$

$$8\pi P_r = \frac{(f+g)^2}{F^2}\left[\frac{\dot{F}^2}{F^2} + \frac{2\dot{F}\dot{f}}{F(f+g)}\right.$$
$$\left. - \frac{3\dot{f}^2}{(f+g)^2} + \frac{2\ddot{f}}{f+g} - \frac{2\ddot{F}}{F}\right]$$
$$+ 3g'^2 - \frac{\alpha^2(f+g)^2}{F^2}, \tag{31}$$

$$8\pi P_\perp = \frac{(f+g)^2}{F^2}\left[\frac{\dot{F}^2}{F^2} - \frac{3\dot{f}^2}{(f+g)^2} + \frac{2\ddot{f}}{f+g} - \frac{\ddot{F}}{F}\right]$$
$$+ 3g'^2 - 2g''(f+g). \tag{32}$$

Next, the matching conditions (22) and (23) on the surface $r = r_\Sigma = $ *constant* read

$$\dot{R}_\Sigma^2 + \alpha^2(R_\Sigma^2 - 2MR_\Sigma - \omega R_\Sigma^4) = 0, \tag{33}$$

and

$$2\ddot{R}_\Sigma R_\Sigma - \dot{R}_\Sigma^2 - \alpha^2(3\omega R_\Sigma^4 - R_\Sigma^2) = 0, \tag{34}$$

with $\omega \equiv g'(r_\Sigma)^2$.

Since (33) is just the first integral of (34), boundary conditions provide only one additional equation.

In order to specify a solution we still need to impose two additional conditions.

One of these conditions will be the quasi–homologous condition which implies because of (20) that the fluid is shear–free ($\sigma = 0$), implying in its turn

$$\frac{\dot{B}}{B} = \frac{\dot{R}}{R} \Rightarrow F(t) = Const. \equiv F_0. \tag{35}$$

Thus the metric functions become

$$A = \frac{F_0}{f(t) + g(r)}, \quad B = \frac{1}{f(t) + g(r)}, \quad R = \frac{F_0}{\alpha\left[f(t) + g(r)\right]}. \tag{36}$$

In order to determine $g(r)$, we shall further impose the vanishing complexity factor condition ($Y_{TF} = 0$), producing

$$g(r) = c_1 r + c_2, \tag{37}$$

with $c_1 \equiv -\sqrt{\omega}$ (we have chosen the minus sign in the definition of $c_1$ to satisfy the condition $R' > 0$), and $c_2$ is another integration constant.

Then from the condition $M = 0$, we obtain a solution to (33) which reads (see [8] for details)

$$R_\Sigma^{(I)} = \frac{1}{\sqrt{\omega} \cos[\alpha(t - t_0)]}. \tag{38}$$

Finally using (38) in (36) we obtain the explicit form of $f(t)$ for this solution

$$f^{(I)}(t) = \frac{1}{\alpha} F_0 \sqrt{\omega} \cos[\alpha(t - t_0)] + \sqrt{\omega} r_\Sigma - c_2. \tag{39}$$

Thus, the corresponding physical variables read

$$\begin{aligned} 8\pi\mu &= -2\omega \cos^2[\alpha(t - t_0)] \\ &+ \omega\left(1 - \frac{r}{r_\Sigma}\right)\left[1 - \frac{r}{r_\Sigma} + 2\cos[\alpha(t - t_0)]\right], \end{aligned} \tag{40}$$

$$8\pi P_r = -\omega\left(1 - \frac{r}{r_\Sigma}\right)\left[4\cos[\alpha(t - t_0)] + 1 - \frac{r}{r_\Sigma}\right], \tag{41}$$

$$8\pi P_\perp = \omega \cos^2[\alpha(t - t_0)] - 2\omega\left(1 - \frac{r}{r_\Sigma}\right)\cos[\alpha(t - t_0)]. \tag{42}$$

The graphics of these physical variables for the solution I are given in Figure 1.

Solution I describes an expanding sphere, whose initial boundary areal radius grows from $1/\sqrt{\omega}$ at $\alpha(t - t_0) = 0$, to infinity as $\alpha(t - t_0) \to \pi/2$, and a contracting sphere whose boundary areal radius decreases from infinity at $\alpha(t - t_0) = 3\pi/2$ to $1/\sqrt{\omega}$ at $\alpha(t - t_0) = 2\pi$. This picture repeating each time interval $\alpha(t - t_0) = 2n\pi$, for any positive real integer $n$.

In order to determine the regions of the fluid distribution where the energy-density is negative (required to have a vanishing total mass) we shall write the condition $\mu = 0$ from (40) in the form

$$z^2 - 2z(1 + \cos y) + 2\cos y(1 - \cos y) + 1 = 0, \tag{43}$$

whose solution reads

$$z = 1 + \cos y - \sqrt{3}\cos y, \tag{44}$$

with $z \equiv \frac{r}{r_\Sigma}$ and $y \equiv \alpha(t - t_0)$.

The curve in Figure 2 is formed by all points in the plane $z, y$ where the energy-density vanishes. The curve divides the plane in two regions, corresponding to negative and positive values of $\mu$. As is apparent from the graphic of $\mu$ in Figure 1, these regions are denoted by A and B respectively, in Figure 2.

### 3.1.2 Solution II: $q = 0; \chi^\alpha || V^\alpha$

We shall next analyze the case when the CKV is parallel to the four–velocity vector in the absence of dissipation. In this case the equation (27) produces

$$A = Bh(r), \qquad R = Br, \qquad \chi^0 = 1, \qquad \psi = \frac{\dot{B}}{B}, \tag{45}$$

where $h(r)$ is an arbitrary function of its argument and $\chi^1 = 0$. It is worth noticing that in this case the fluid is necessarily shear–free, implying thereby that it evolves in QH regime.

Thus the line element may be written as

$$ds^2 = B^2\left[-h^2(r)dt^2 + dr^2 + r^2(d\theta^2 + \sin^2\theta d\phi^2)\right]. \tag{46}$$

Next, using (45) and the field equations, the condition $q = 0$ reads (see [8] for details)

$$\frac{\dot{R}'}{R} - 2\frac{\dot{R}}{R}\frac{R'}{R} - \frac{\dot{R}}{R}\left(\frac{h'}{h} - \frac{1}{r}\right) = 0, \tag{47}$$

whose solution is

$$R = \frac{r}{h(r)\left[f(t) + g(r)\right]}, \tag{48}$$

implying

$$B = \frac{1}{h(r)\left[f(t) + g(r)\right]}, \quad A = \frac{1}{f(t) + g(r)}, \tag{49}$$

where $g$ and $f$ are two arbitrary functions of their argument.

Thus the metric is defined up to three arbitrary functions $(g(r), f(t), h(r))$.

The function $f(t)$ will be obtained from the junction conditions (22), (23).

Indeed, evaluating the mass function at the boundary surface $\Sigma$ we obtain from (22) and (48)

$$\dot{R}_\Sigma^2 = \alpha^2 R_\Sigma^4 \left[\epsilon - V(R_\Sigma)\right], \tag{50}$$

where

$$\alpha^2 \equiv \frac{h_\Sigma^2}{r_\Sigma^2}, \qquad \epsilon \equiv (g')_\Sigma^2 h_\Sigma^2, \tag{51}$$

and

$$V(R_\Sigma) = \frac{2\sqrt{\epsilon}}{R_\Sigma}(1 - a_1) + \frac{a_1}{R_\Sigma^2}(2 - a_1) - \frac{2M}{R_\Sigma^3}, \tag{52}$$

with $a_1 \equiv \frac{h'_\Sigma r_\Sigma}{h_\Sigma}$.

On the other hand, from (23), using (48) we obtain

$$2\ddot{R}_\Sigma R_\Sigma - \dot{R}_\Sigma^2 - 3\epsilon\alpha^2 R_\Sigma^4 - 4\alpha^2 \sqrt{\epsilon} R_\Sigma^3 (a_1 - 1)$$
$$-\alpha^2 R_\Sigma^2 a_1 (a_1 - 2) = 0. \tag{53}$$

Thus assuming $M = 0$, equation (50) becomes

$$\dot{R}_\Sigma^2 = \alpha^2 R_\Sigma^4 \left[\epsilon - \frac{2\sqrt{\epsilon}(1 - a_1)}{R_\Sigma} - \frac{a_1(2 - a_1)}{R_\Sigma^2}\right]. \tag{54}$$

Solutions to the above equation in terms of elementary functions may be obtained by assuming $a_1 = 1$, in which case a possible solution to (54) is

$$R_\Sigma^{(II)} = \frac{1}{\sqrt{\epsilon}\cos[\alpha(t - t_0)]}, \tag{55}$$

which exhibits the same time dependence as in solution $I$.

Thus, as in the previous model, solution $II$ describes an expanding sphere, whose initial boundary areal radius grows from $1/\sqrt{\epsilon}$ at $\alpha(t - t_0) = 0$, to infinity as $\alpha(t - t_0) \to \pi/2$, and a contracting sphere whose boundary areal radius decreases from infinity at $\alpha(t-t_0) = 3\pi/2$ to $1/\sqrt{\epsilon}$ at $\alpha(t-t_0) = 2\pi$. This picture repeating each time interval $\alpha(t-t_0) = 2n\pi$, for any positive real integer $n$.

Imposing further the vanishing complexity factor condition, then functions $h(r), g(r)$ are given by

$$h(r) = c_6 r, \qquad g = c_7 \ln r + c_5. \tag{56}$$

The physical variables corresponding to this solution read

$$\begin{aligned} 8\pi\mu &= -2\epsilon \cos^2[\alpha(t-t_0)] + \\ &+ \epsilon \ln \frac{r}{r_\Sigma} \left[ 2\cos[\alpha(t-t_0)] + \ln \frac{r}{r_\Sigma} \right], \end{aligned} \tag{57}$$

$$8\pi P_r = -\epsilon \ln \frac{r}{r_\Sigma} \left[ 4\cos[\alpha(t-t_0)] + \ln \frac{r}{r_\Sigma} \right], \tag{58}$$

$$8\pi P_\perp = \epsilon \cos^2[\alpha(t-t_0)] - 2\epsilon \ln \frac{r}{r_\Sigma} \cos[\alpha(t-t_0)]. \tag{59}$$

The behavior of these physical variables is depicted in Figure 3.

From the definition of the mass function (11), using (48), (49) (55) and (56), the condition $m(r_\Sigma) = 0$ implies

$$\epsilon \cos^2[\alpha(t-t_0)] \left(1 - \frac{\alpha^2}{c_6^2}\right) + \frac{\epsilon \alpha^2}{c_6^2} - c_7^2 c_6^2 = 0. \tag{60}$$

The above equation is satisfied for any value of $t$ if $\alpha^2 = c_6^2$ and $\epsilon = c_7^2 \alpha^2$, which, as expected, are the same relationships which follow from (51) and (56).

In oder to determine the regions of the fluid distribution where the energy-density is negative (required to have a vanishing total mass) we shall write the condition $\mu = 0$ from (57) in the form

$$\ln z \left(2\cos y + \ln z\right) - 2\cos^2 y = 0, \tag{61}$$

whose solution reads

$$z = e^{-(1+\sqrt{3})\cos y}, \tag{62}$$

where $z \equiv \frac{r}{r_\Sigma}$, $y \equiv \alpha(t-t_0)$.

The graphic of $z$ as function of $y$ is plotted in Figure 4. The curve contains all the points of the plane $z, y$ where the energy-density vanishes, and divides the plane in two regions ($A$ and $B$) corresponding to negative and positive energy-density respectively.

## 3.2 Expansion–free models

We shall now present models for which the expansion scalar vanishes.

We recall that under such a condition the line element may be written as (see [24] for details)

$$ds^2 = -\left\{\frac{R^2 \dot{R}}{\alpha} \exp\left[-4\pi \int qAB \frac{R}{\dot{R}} dr\right]\right\}^2 dt^2$$
$$+ \frac{\alpha^2}{R^4} dr^2 + R^2(d\theta^2 + \sin^2\theta d\phi^2), \tag{63}$$

where $\alpha$ is a unit constant with dimensions $[r^2]$.

Besides, as is known, the expansion-free models are endowed with an internal vacuum cavity surrounding the center, accordingly the center of symmetry is not filled with fluid.

The solutions deployed below satisfy the vanishing complexity factor condition and the quasi-homologous evolution.

### 3.2.1 Solution III: $Y_{TF} = 0$, $A = A(r)$, $R = R_1(t)R_2(r)$

For this model, we shall complement the expansion–free condition with the vanishing complexity factor condition $Y_{TF} = 0$, and we shall assume that $A$ only depends on the radial coordinate, and $R$ is a separable function (i.e., $R = R_1(t)R_2(r)$).

From all these conditions the general form of the metric variables read (see [24] for details)

$$R_1(t) = \frac{\nu_0}{t + \nu_1}, \tag{64}$$

$$R_2(r) = \nu_2 A^{\nu_3 - 1}, \tag{65}$$

$$A = \nu_4 \left(\nu_3 r + \nu_5\right)^{\frac{1}{\nu_3}}, \tag{66}$$

where $\nu_0$ and $\nu_3$ are dimensionless constants and $\nu_1, \nu_2, \nu_4$ and $\nu_5$ are constants with dimensions $[r], [r^2], [1/(r^{1/\nu_3})]$ and $[r]$, respectively.

Let us choose for our model

$$\nu_0 = 1, \nu_1 = \nu_5 = 0; \nu_3 = 2; \nu_2 = \frac{1}{\nu_4^4}. \tag{67}$$

Then the expression for the mass function evaluated at the boundary surface and the areal radius of the boundary surface become

$$m(r_\Sigma) = \frac{\alpha^{1/3} r_\Sigma^{1/3}}{\sqrt{2} t^*} \left( \frac{1}{t^{*4}} + 1 - \frac{2}{t^{*6}} \right), \tag{68}$$

and

$$R(r_\Sigma) = \frac{\alpha^{1/3} r_\Sigma^{1/3} \sqrt{2}}{t^*}, \tag{69}$$

where $t^* = \nu_4^2 t$ and we have put $\frac{r_\Sigma}{\alpha^2 \nu_4^6} = 1$.

From (68) it follows that the total mass vanishes at $t^* = 1$.

The physical variables for this model read

$$8\pi\mu = \frac{1}{\alpha^{2/3} r_\Sigma^{2/3}} \left( \frac{t^{*2}}{2x} - \frac{3}{t^{*4}} - \frac{3}{2xt^{*2}} \right), \tag{70}$$

$$4\pi q = -\frac{1}{\alpha^{2/3} r_\Sigma^{2/3}} \left( \frac{\sqrt{2}}{t^{*3} \sqrt{x}} \right), \tag{71}$$

$$8\pi P_r = -\frac{1}{\alpha^{2/3} r_\Sigma^{2/3}} \left( \frac{5}{2xt^{*2}} + \frac{t^{*2}}{2x} - \frac{3}{t^{*4}} \right), \tag{72}$$

$$8\pi P_\perp = \frac{1}{\alpha^{2/3} r_\Sigma^{2/3}} \left( \frac{3}{t^{*4}} - \frac{1}{xt^{*2}} \right), \tag{73}$$

$$T = \frac{\alpha^{1/3}}{\sqrt{2x} r_\Sigma^{2/3}} \left[ T_0(t^*) + \frac{3\tau}{2\pi\kappa t^{*2} \alpha \sqrt{2x}} + \frac{r_\Sigma^{1/3} \ln(2xr_\Sigma)}{4\pi\kappa \alpha^{2/3} t^*} \right], \tag{74}$$

where the expression above for the temperature has been obtained using the truncated transport equation (26), and $x \equiv \frac{r}{r_\Sigma}$.

The "effective energy–density" appearing in the definition of the mass function (15) $\left( \mu + q\frac{U}{E} \right)$ evaluated at $t^* = 1$ reads

$$8\pi \alpha^{2/3} r_\Sigma^{2/3} \left( \mu + \frac{qU}{E} \right) = \frac{1}{x} - 3. \tag{75}$$

Figure 5 and Figure 6 depict the behavior of physical variables, the radial distribution of the "effective energy–density" at $t^* = 1$, when the total mass vanishes, and the evolution of the total mass.

The model represents a contracting sphere with initial negative mass absorbing energy through the boundary surface. At $t^* = 1$ the total mass vanishes becoming positive afterward.

It is worth mentioning that although the total mass tends to zero as $t \to \infty$, the fluid distribution does not characterize a ghost star in that limit, since in such a case the total mass tends to zero due to the fact that the integrand in (15) tends to zero as $t \to \infty$, and not because of change of sign of the effective energy-density as is the case for a ghost star.

### 3.2.2 $A = 1$, $Y_{TF} = 0$

For this model we shall assume that the fluid is geodesic, meaning

$$A(t, r) = 1, \tag{76}$$

the above condition together with the expansion-free condition

$$B(t, r) = \frac{\alpha}{R^2}, \tag{77}$$

plus the condition $Y_{TF} = 0$, produces

$$R = \frac{\alpha}{[t + \alpha b_2(r)]}, \qquad B = \frac{[t + \alpha b_2(r)]^2}{\alpha}, \tag{78}$$

where $b_2$ is an arbitrary function of its argument with dimensions $[1/r]$.

For our model we shall choose

$$b_2 = -\frac{\sqrt{6}r}{\alpha}, \qquad \alpha = 6r_\Sigma^2, \tag{79}$$

producing

$$R = \frac{6^{(1/2)} r_\Sigma}{t^* - x}, \qquad B = (t^* - x)^2, \tag{80}$$

where

$$t^* = \frac{t}{\sqrt{6} r_\Sigma}, \qquad r = x r_\Sigma, \tag{81}$$

and the mass function becomes

$$m = \frac{3 r_\Sigma}{\sqrt{6}(t^* - x)^9} \left[(t^* - x)^8 + (t^* - x)^4 - 6\right]. \tag{82}$$

From (82) evaluated at $r = r_\Sigma$ we see that $m(r_\Sigma) = 0$ at $t^* = (2)^{1/4} + 1 \approx 2.19$, being negative before that time and becoming positive afterward.

For this metric, the physical variables and the shear read as follows:

$$8\pi\mu = -\frac{1}{2r_\Sigma^2 (t^* - x)^2} - \frac{9}{r_\Sigma^2 (t^* - x)^6} + \frac{(t^* - x)^2}{6r_\Sigma^2}, \tag{83}$$

$$4\pi q = -\frac{2\sqrt{6}}{3r_\Sigma^2 (t^* - x)^4}, \tag{84}$$

$$8\pi P_r = -\frac{5}{6r_\Sigma^2 (t^* - x)^2} + \frac{1}{r_\Sigma^2 (t^* - x)^6} - \frac{(t^* - x)^2}{6r_\Sigma^2}, \tag{85}$$

$$8\pi P_\perp = -\frac{1}{3r_\Sigma^2 (t^* - x)^2} + \frac{4}{r_\Sigma^2 (t^* - x)^6}, \tag{86}$$

$$\sigma = \frac{3}{\sqrt{6} r_\Sigma (t^* - x)}, \tag{87}$$

$$T = T_0(t) + \frac{1}{\pi\kappa\sqrt{6}r_\Sigma(t^* - x)}\left[1 - \frac{2\tau}{\sqrt{6}r_\Sigma(t^* - x)}\right], \tag{88}$$

where, as in the previous case, the temperature has been calculated using the truncated transport equation (26).

Figure 7 and Figure 8 depict the behavior of physical variables, the radial distribution of the "effective energy–density" at $t^* = (2)^{1/4} + 1$, when the total mass vanishes, and the evolution of the total mass.

The model represents a contracting sphere with initial negative mass absorbing energy through the boundary surface. At $t^* = (2)^{1/4} + 1$ the total mass vanishes becoming positive afterward. From the above it follows that

$$8\pi r_\Sigma^2 \left(\mu + q\frac{U}{E}\right) = \frac{5}{6(t^* - x)^2} - \frac{9}{(t^* - x)^6} + \frac{(t^* - x)^2}{6}. \tag{89}$$

Evaluating the above expression at $t^* = (2)^{1/4} + 1$ we see that it changes of sign within the fluid distribution, thereby explaining the vanishing of the total mass at $t^* = (2)^{1/4} + 1$(see Figure 8).

# 4   Discussion

We have presented a set of solutions of fluid spheres whose evolution involves ghost stars.

The first two solutions represent the adiabatic evolution of a ghost star, they admit a CKV which may be either orthogonal or parallel to the four-velocity vector. These solutions ($I$ and $II$) describe either an expanding sphere, with an initial boundary areal radius growing from some finite value to infinity, or a contracting sphere whose boundary areal radius decreases from infinity to some finite final value. In both cases $m(r_\Sigma) = 0$ at all times and the energy–density is negative in some regions of the fluid distributions (see Figures 2 and 4). The full description of these models is provided by equations (38-42), and (55-60) for models $I$ and $II$ respectively. Their behavior is depicted in Figures 1, 2, 3, 4.

In both cases the vanishing complexity factor condition applies and the solutions match smoothly to the Minkowski space-time on the boundary surface of the fluid distribution.

The second couple of solutions satisfies the expansion-free condition and the vanishing complexity factor condition. In one case (solution $III$), these last conditions are complemented with the assumption that $A = A(r)$ and $R$ is a separable function. In the last case (solution $IV$) we assume the fluid to be geodesic.

The full description of model $III$ is provided by Equations (64–75) and illustrated in Figures 5 and 6. It describes a collapsing fluid whose total mass evolves from negative values to positive ones by absorbing radiation. At some point of the evolution ($t^* = 1$) the total mass vanishes becoming positive afterward. At $t^* = 1$ the effective energy–density is negative in some regions of the fluid distributions as shown in Figure 6.

This rather unusual scenario (compact object absorbing radiation), has been invoked in the past to explain the origin of gas in quasars [34]. A semi-numerical example for such a model is described in [35].

Finally, the last model is geodesic (solution IV) and satisfies, besides the expansion–free condition, the vanishing complexity factor condition. It is described by Equations (80–88). This solution depicts a collapsing fluid for which as $t \to \infty$, the energy density and the radial pressure diverge and satisfy the equation of the state $\mu = -P_r$, whereas the heat flux vector and the tangential pressure vanish, and the temperature tends toward $T_0$. As it happens in Solution $III$, at some point of its evolution $t^* = (2)^{1/4} + 1$ the total mass vanishes, and as expected the effective energy–density is negative in some regions of the fluid distributions as shown in Figure 8.

Unlike solutions $I, II$, solutions $III, IV$ do not satisfy Darmois conditions on the boundary surface, implying that these surfaces are thin shells.

It is worth recalling that the very existence of ghost stars relies on the presence of negative energy–density (the effective energy–density in the non–adiabatic case). Negative energy-density (mass) is a condition which has been shown to be likely to appear in scenarios where quantum effects are relevant (see [36, 37, 38, 39, 40, 41] and references therein).

An issue requiring much more research work concerns the possibility to observe a ghost star. We have in mind either a "permanent" ghost star as the case described by solutions $I$ and $II$, or a compact object attaining momentarily the ghost star status, as the two models described by solutions $III$ and $IV$.

At present we contemplate three possible ways to establish (or dismiss) the very existence of a ghost star. On the one hand, by observing the shadow of such objects following the line of research open by the Event Horizon Telescope (EHT) Collaboration (see [42, 43, 44, 45] and references therein). On the other hand the formation of a ghost star, even if for a short time interval, involves radiating processes whose observation could help to identify a ghost star. Finally a ghost star exhibits null gravitational surface redshift, a fact that could be used to its identification. Models as $III$ and $IV$ seem to be particularly suitable for this purpose.

In relationship with this last point it would be very helpful to find an evolving model , with a positive energy flux at the boundary surface, leading asymptotically to a ghost star. Unfortunately, neither of the solutions presented here satisfy such a condition.

## 5 Einstein equations

Einstein's field equations for the interior spacetime (1) are given by

$$G_{\alpha\beta} = 8\pi T_{\alpha\beta}, \tag{90}$$

and its non zero components read

$$8\pi T_{00} = 8\pi\mu A^2 = \left(2\frac{\dot{B}}{B} + \frac{\dot{R}}{R}\right)\frac{\dot{R}}{R} - \left(\frac{A}{B}\right)^2\left[2\frac{R''}{R} + \left(\frac{R'}{R}\right)^2 - 2\frac{B'}{B}\frac{R'}{R} - \left(\frac{B}{R}\right)^2\right], \tag{91}$$

$$8\pi T_{01} = -8\pi qAB = -2\left(\frac{\dot{R}'}{R} - \frac{\dot{B}}{B}\frac{R'}{R} - \frac{\dot{R}}{R}\frac{A'}{A}\right), \quad (92)$$

$$8\pi T_{11} = 8\pi P_r B^2 = -\left(\frac{B}{A}\right)^2\left[2\frac{\ddot{R}}{R} - \left(2\frac{\dot{A}}{A} - \frac{\dot{R}}{R}\right)\frac{\dot{R}}{R}\right] + \left(2\frac{A'}{A} + \frac{R'}{R}\right)\frac{R'}{R} - \left(\frac{B}{R}\right)^2, \quad (93)$$

$$8\pi T_{22} = \frac{8\pi}{\sin^2\theta}T_{33} = 8\pi P_\perp R^2 = -\left(\frac{R}{A}\right)^2\left[\frac{\ddot{B}}{B} + \frac{\ddot{R}}{R} - \frac{\dot{A}}{A}\left(\frac{\dot{B}}{B} + \frac{\dot{R}}{R}\right) + \frac{\dot{B}}{B}\frac{\dot{R}}{R}\right]$$
$$+ \left(\frac{R}{B}\right)^2\left[\frac{A''}{A} + \frac{R''}{R} - \frac{A'}{A}\frac{B'}{B} + \left(\frac{A'}{A} - \frac{B'}{B}\right)\frac{R'}{R}\right]. \quad (94)$$

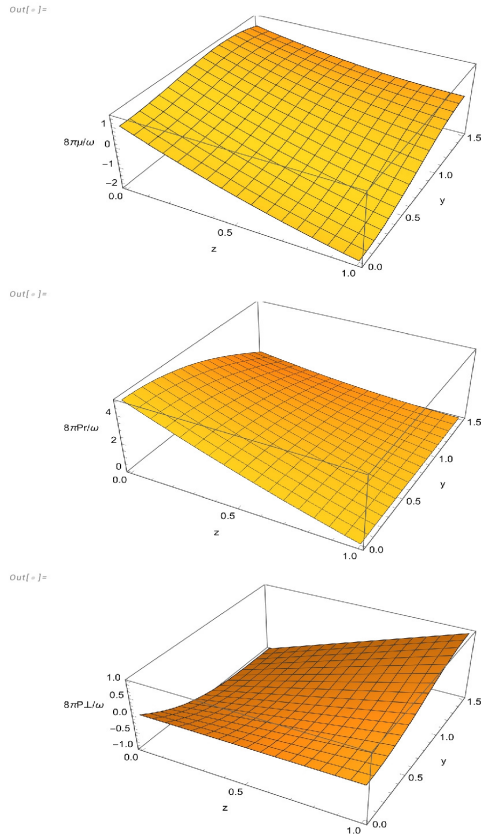

Figure 1: $8\pi\mu/\omega$, $8\pi P_r/\omega$ and $8\pi P_\perp/\omega$, as functions of $y \equiv \frac{\alpha}{2}(t - t_0)$ and $z \equiv \frac{r}{r_\Sigma}$ for solution I.

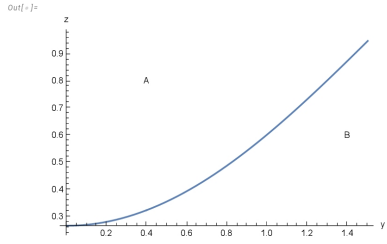

Figure 2: $z \equiv \frac{r}{r_\Sigma}$ as function of $y \equiv \alpha(t-t_0)$ for the condition $\mu = 0$. Regions $A$ and $B$ correspond to negative and positive values of $\mu$ respectively, for solution I.

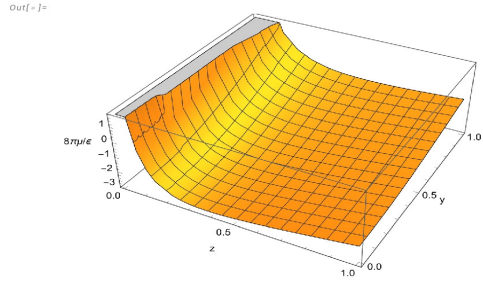
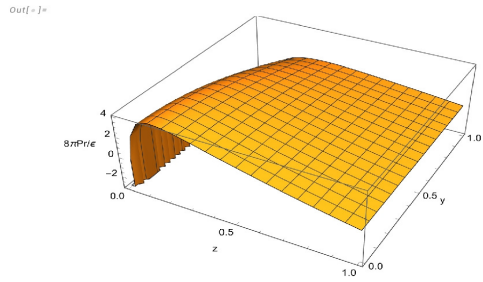
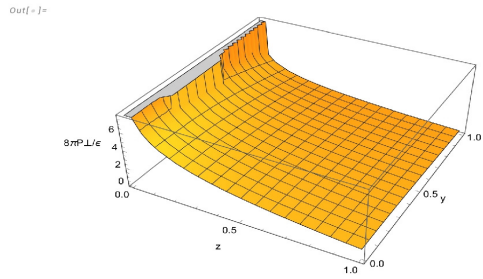

Figure 3: $8\pi\mu/\epsilon$, $8\pi P_r/\epsilon$ and $8\pi P_\perp/\epsilon$, as functions of $y \equiv \frac{\alpha}{2}(t-t_0)$ and $z \equiv \frac{r}{r_\Sigma}$ for solution II.

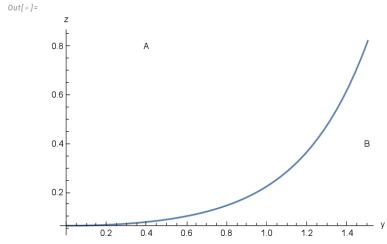

Figure 4: $z \equiv \frac{r}{r_\Sigma}$ as function of $y \equiv \alpha(t-t_0)$ for the condition $\mu = 0$. Regions $A$ and $B$ correspond to negative and positive values of $\mu$ respectively, for solution II.

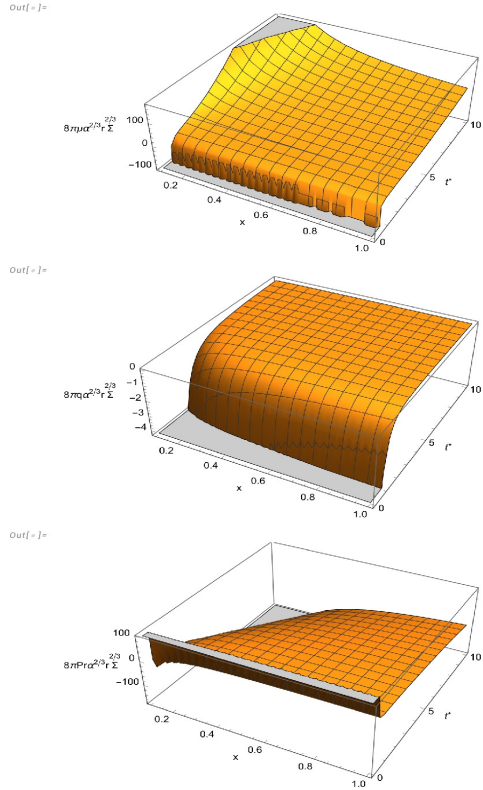

Figure 5: $8\pi\mu\alpha^{2/3}r_\Sigma^{2/3}$, $8\pi q\alpha^{2/3}r_\Sigma^{2/3}$ and $8\pi P_r\alpha^{2/3}r_\Sigma^{2/3}$, as functions of $t^* \equiv \frac{tr_\Sigma^{1/3}}{\alpha^{2/3}}$ and $x \equiv \frac{r}{r_\Sigma}$ for solution III.

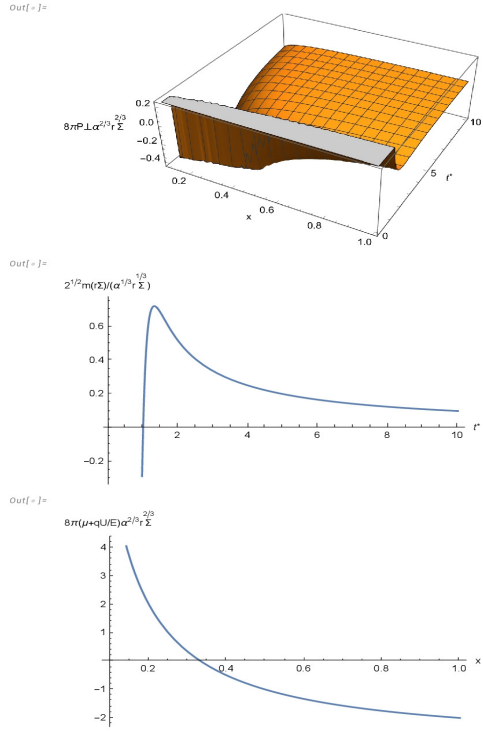

Figure 6: $8\pi P_\perp \alpha^{2/3} r_\Sigma^{2/3}$ as function of $t^* \equiv \frac{t r_\Sigma^{1/3}}{\alpha^{2/3}}$ and $x \equiv \frac{r}{r_\Sigma}$; $2^{1/2} m(r\Sigma)/\alpha^{1/3} r_\Sigma^{1/3}$ as function of $t^*$; $8\pi(\mu + q\frac{U}{E})\alpha^{2/3} r_\Sigma^{2/3}$, evaluated at $t^* = 1$, as function of $x$, for solution III.

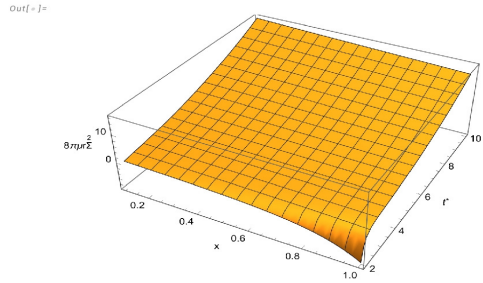

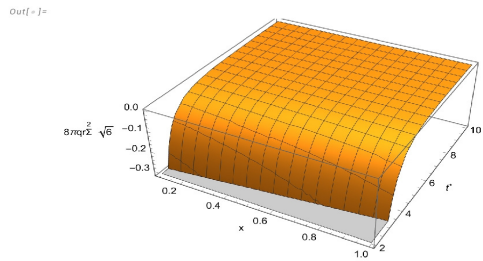

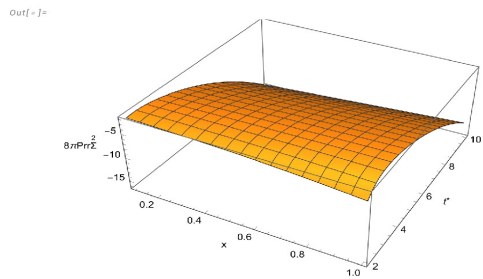

Figure 7: $8\pi\mu r_\Sigma^2$, $8\pi q r_\Sigma^2$ and $8\pi P_r r_\Sigma^2$, as functions of $t^* \equiv \frac{t}{r_\Sigma \sqrt{6}}$ and $x \equiv \frac{r}{r_\Sigma}$ for solution IV.

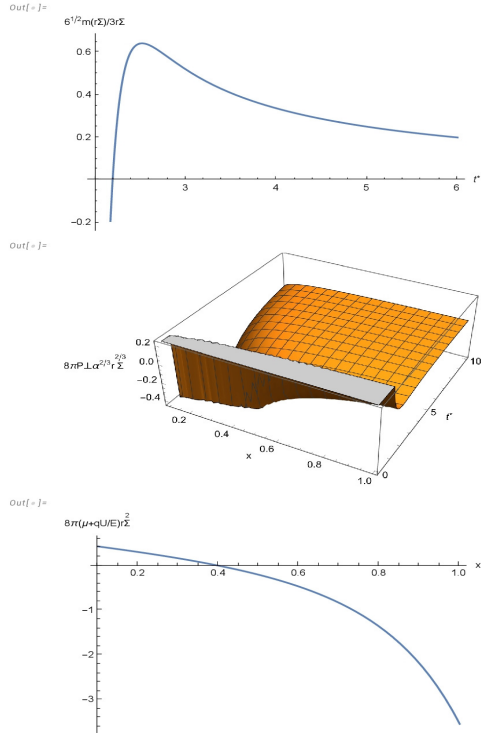

Figure 8: $8\pi P_\perp r_\Sigma^2$ as function of $t^* \equiv \frac{t}{r_\Sigma \sqrt{6}}$ and $x \equiv \frac{r}{r_\Sigma}$; $\frac{6^{1/2}m(r_\Sigma)}{3r_\Sigma}$ as function of $t^*$; $8\pi r_\Sigma^2(\mu + q\frac{U}{E})$, evaluated at $t^* = 1$, as function of $x$, for solution IV.